\documentclass[floatfix,aps,prl,twocolumn,superscriptaddress]{revtex4}
\usepackage{graphicx}
\usepackage[]{epsfig}
\usepackage{times,amsmath,amssymb}
\begin{document}

\title{Fast probe of local electronic states in nanostructures utilizing a single-lead quantum dot}

\author{Tomohiro Otsuka}
\email[]{tomohiro.otsuka@riken.jp}
\affiliation{Center for Emergent Matter Science, RIKEN, 2-1 Hirosawa, Wako, Saitama 351-0198, Japan}
\affiliation{Department of Applied Physics, University of Tokyo, 7-3-1 Hongo, Bunkyo, Tokyo 113-8656, Japan}

\author{Shinichi Amaha}
\affiliation{Center for Emergent Matter Science, RIKEN, 2-1 Hirosawa, Wako, Saitama 351-0198, Japan}

\author{Takashi Nakajima}
\affiliation{Center for Emergent Matter Science, RIKEN, 2-1 Hirosawa, Wako, Saitama 351-0198, Japan}
\affiliation{Department of Applied Physics, University of Tokyo, 7-3-1 Hongo, Bunkyo, Tokyo 113-8656, Japan}

\author{Matthieu R. Delbecq}
\affiliation{Center for Emergent Matter Science, RIKEN, 2-1 Hirosawa, Wako, Saitama 351-0198, Japan}
\affiliation{Department of Applied Physics, University of Tokyo, 7-3-1 Hongo, Bunkyo, Tokyo 113-8656, Japan}

\author{Jun Yoneda}
\affiliation{Center for Emergent Matter Science, RIKEN, 2-1 Hirosawa, Wako, Saitama 351-0198, Japan}
\affiliation{Department of Applied Physics, University of Tokyo, 7-3-1 Hongo, Bunkyo, Tokyo 113-8656, Japan}

\author{Kenta Takeda}
\affiliation{Center for Emergent Matter Science, RIKEN, 2-1 Hirosawa, Wako, Saitama 351-0198, Japan}
\affiliation{Department of Applied Physics, University of Tokyo, 7-3-1 Hongo, Bunkyo, Tokyo 113-8656, Japan}

\author{Retsu Sugawara}
\affiliation{Center for Emergent Matter Science, RIKEN, 2-1 Hirosawa, Wako, Saitama 351-0198, Japan}
\affiliation{Department of Applied Physics, University of Tokyo, 7-3-1 Hongo, Bunkyo, Tokyo 113-8656, Japan}

\author{Giles Allison}
\affiliation{Center for Emergent Matter Science, RIKEN, 2-1 Hirosawa, Wako, Saitama 351-0198, Japan}
\affiliation{Department of Applied Physics, University of Tokyo, 7-3-1 Hongo, Bunkyo, Tokyo 113-8656, Japan}

\author{Arne Ludwig}
\affiliation{Angewandte Festk\"orperphysik, Ruhr-Universit\"at Bochum, D-44780 Bochum, Germany}

\author{Andreas D. Wieck}
\affiliation{Angewandte Festk\"orperphysik, Ruhr-Universit\"at Bochum, D-44780 Bochum, Germany}

\author{Seigo Tarucha}%
\affiliation{Center for Emergent Matter Science, RIKEN, 2-1 Hirosawa, Wako, Saitama 351-0198, Japan}
\affiliation{Department of Applied Physics, University of Tokyo, 7-3-1 Hongo, Bunkyo, Tokyo 113-8656, Japan}
\affiliation{Quantum-Phase Electronics Center, University of Tokyo, 7-3-1 Hongo, Bunkyo, Tokyo 113-8656, Japan}
\affiliation{Institute for Nano Quantum Information Electronics, University of Tokyo, 4-6-1 Komaba, Meguro, Tokyo 153-8505, Japan}

\date{\today}
\begin{abstract}
Transport measurements are powerful tools to probe electronic properties of solid-state materials. 
To access properties of local electronic states in nanostructures, such as local density of states, electronic distribution and so on, micro-probes utilizing artificial nanostructures have been invented to perform measurements in addition to those with conventional macroscopic electronic reservoirs.
Here we demonstrate a new kind of micro-probe: a fast single-lead quantum dot probe, which utilizes a quantum dot coupled only to the target structure through a tunneling barrier and fast charge readout by RF reflectometry.
The probe can directly access the local electronic states with wide bandwidth.
The probe can also access more electronic states, not just those around the Fermi level, and the operations are robust against bias voltages and temperatures.
\end{abstract}

\pacs{73.22.-f, 73.23.-b, 73.63.Kv, 85.35.-p}
\maketitle

%%% Introduction %%%

New kinds of structures based on solid-state nanostructures have been proposed to realize functional devices.
For example, spintronics devices utilizing spin effects~\cite{2001WolfSci, 2004ZuticRMP} and quantum information-processing devices utilizing quantum effects in nanostructures~\cite{1998LossPRA, 2000NielsenBk, 2010LaddNat} have been proposed and studied intensively.
In these new devices, local electronic states play important roles and their understanding on a microscopic basis is crucial.

Transport measurements are one of the most powerful tools to probe electronic properties of nanostructures.
In conventional transport measurements, macroscopic probes, electronic reservoirs that contain huge ensembles of electrons, are coupled to the target system and the flow of electrons is measured to probe electronic properties~\cite{1997DattaBk}.
Using such macroscopic electronic reservoirs brings some constraints in the measurement of nanostructures: direct access to small local regions is not easy, the transport is limited to only around the Fermi level, the measurement is greatly affected by change of the electronic distribution in the reservoir, for example caused by bias voltages or electron temperatures, and the measurement is usually slow because of the large geometric capacitances of the leads.
To overcome these restrictions is a strong challenge in transport measurements.

One possible solution to this challenge requires microscopic probes utilizing nanostructures instead of macroscopic reservoirs.
The use of semiconductor quantum dots (QDs) in such probes has been demonstrated.
QDs have well defined inner quantum levels that can be controlled by applying voltages on gate electrodes~\cite{1996TaruchaPRL, 1997KouwenhovenScience, 2000CiorgaPRB}.
By measuring the transport through these artificial quantum levels, we directly access local electronic states.
This cannot be realized with conventional macroscopic probes.
For example, measurements of local electronic states, energy relaxation and heat transport in quantum Hall edge states have been demonstrated by utilizing QDs~\cite{2010AltimirasNatPhys, 2010leSueurPRL, 2012VenkatachalamNatPhys}.

In this paper, we realize a new kind of QD probe: a fast single-lead quantum dot (SLQD) probe.
An SLQD is a QD, which couples to a target system through a single tunneling barrier~\cite{2008OtsukaAPL,2009OtsukaPRB, 2010OtsukaPRB, 2012OtsukaPRB}.
We can probe more states, not only those around the Fermi level, with robustness against
change of the electronic distribution because the SLQD is fully isolated from
electronic reservoirs.
Also, we can improve the operation
time of the probe because the geometric capacitance of the SLQD is small and by utilizing RF reflectometry
techniques~\cite{1998SchoelkopfSci, 2007ReillyAPL, 2010BarthelPRB}.

First, we realized the fast SLQD probe and evaluated its operation speed.
Second, we applied the probe to detect local electronic states in another QD, which is used as a controllable target nanostructure in this experiment.
By measuring tunneling events of electrons between the probe SLQD and the target QD, we confirmed the operation of the new probe.
Additionally, we demonstrated its key features. 
Finally, we show that this probe can conduct fast real-time measurements of local electronic states.

\section{Results}
\subsection{Realization of a fast single-lead quantum dot probe}
%%% Device & Charge states %%%

\begin{figure}
\begin{center}
  \includegraphics{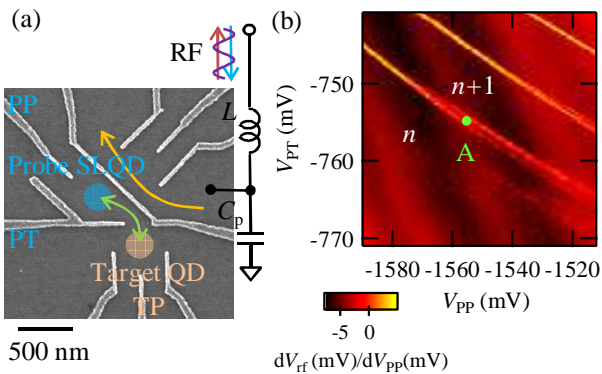}
  \caption{(color online) (a) Scanning electron micrograph of the device and schematic of the measurement setup.
  A probe SLQD and a target QD are formed at the upper left and the lower middle, respectively.
  A QPC charge sensor at the upper right is connected to a resonator for the RF-reflectometry.
  (b) ${\rm d}V_{\rm rf}/{\rm d} V_{\rm PP}$ as a function of $V_{\rm PP}$ and $V_{\rm PT}$.
  Bright lines correspond to the charge transition lines of the probe SLQD.
  }
  \label{Device}
\end{center}
\end{figure}

Figure~\ref{Device}(a) is a scanning electron micrograph of the device.
By applying negative voltages on the gate electrodes, a probe SLQD, a target QD and a QPC charge sensor~\cite{1993FieldPRL, 1998BuksNature, 2002SprinzakPRL, 2003ElzermanPRB} are formed at the upper left, the lower middle and the upper right, respectively.
The QPC charge sensor is connected to an RF resonator formed by an inductor $L$ and a stray capacitance $C_{\rm p}$ (resonance frequency $f_{\rm res}$=211~MHz).
The number of electrons in the probe SLQD $n$ is monitored by the intensity of the reflected RF signal $V_{\rm rf}$.
Another possible option to monitor $n$ will be the dispersive readout using the gates~\cite{2013CollessPRL} of the SLQD, which will simplify the device structure.

First, we formed only the probe SLQD and coupled the probe to the two-dimensional electron gas to check the operation speed of the probe.
Figure~\ref{Device}(b) is a charge state diagram of the SLQD.
We measured $V_{\rm rf}$ as a function of a plunger gate votage $V_{\rm PP}$ and a tunneling gate voltage $V_{\rm PT}$.
To make signals clearer, we plotted numerical differentials of the signals ${\rm d}V_{\rm rf}/{\rm d} V_{\rm PP}$ in Fig.~\ref{Device}(b).
When $n$ changes with change of $V_{\rm PP}$, $V_{\rm rf}$ shows jumps that are observed as bright lines in the figure.
Note that the absolute value of $n$ is not exactly one in this device and is expected around $\sim $10 from the size of the QD and the interval between the charge transition lines~\cite{comment1}.
As $V_{\rm PT}$ is changed to more negative voltages, the charge transition lines become less visible because tunneling rates become small and electron tunneling into the SLQD rarely happens within the time scale of the $V_{\rm PP}$ sweep ($\sim $1ms, $V_{\rm PP}$ was swept by saw-tooth wave voltages with a frequency of 1160~Hz).

%%% Fast readout %%%

\begin{figure}
\begin{center}
  \includegraphics{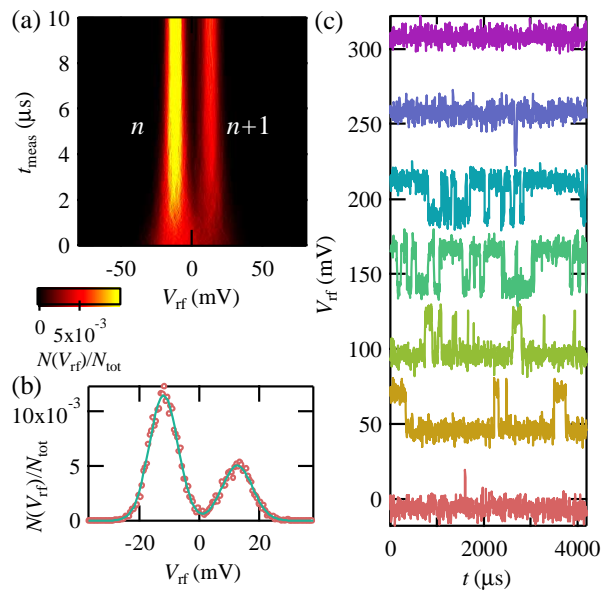}
  \caption{(color online) (a) $N(V_{\rm rf})/N_{\rm tot}$ as a function of $V_{\rm rf}$ and $t_{\rm meas}$.
  With increase of $t_{\rm meas}$, we can distinguish two peaks originating from the $n$ and $n+1$ charge states.
  The left (right) peak corresponds to the $n$ ($n+1$) state.
  (b) $N(V_{\rm rf})/N_{\rm tot}$ as a function of $V_{\rm rf}$ at $t_{\rm meas}$=5$\mu $s.
  We can clearly distinguish the double peaks.
  The trace shows a result of fitting with a double Gaussian.
  (c) $V_{\rm rf}$ as a function of $t$ with changing $V_{\rm PP}$ around point A in Fig.~\ref{Device}(b).
  The charge transition happens around the charge transition lines and observed as jumps of $V_{\rm rf}$ in real-time.
  }
  \label{FastMeas}
\end{center}
\end{figure}

We evaluated the measurement time that is required to resolve a single electron charge in this probe SLQD.
Figure~\ref{FastMeas}(a) shows the number of events $N(V_{\rm rf})/N_{\rm tot}$ as a function of $V_{\rm rf}$ and measurement time $t_{\rm meas}$.
This histogram is produced as a result of $N_{\rm tot}$=8192 repetitions of the measurement.
The gate condition of the probe SLQD is fixed at point A in Fig.~\ref{Device}(b) on a charge transition line.
At this point, the charge state changes between $n$ and $n+1$ states in a time scale of several hundreds of $\mu $s.
In our measurement setup, a reflected RF signal is demodulated, digitized and integrated on $t_{\rm meas}$ to produce $V_{\rm rf}$.
With the increase of $t_{\rm meas}$, it becomes possible to distinguish two peaks originating from $n$ and $n+1$ charge states.
The left (right) peak corresponds to the $n$ ($n+1$) state.
Figure~\ref{FastMeas}(b) is a histogram at $t_{\rm meas}$=5$\mu $s.
The two peaks are well fitted by a double gaussian. We can resolve the $n$ and $n+1$ charge states with fidelity exceeding 99\%.
This value of $t_{\rm meas}$ is much shorter than values in previous experiments with conventional slow electronics ($\sim $ms)~\cite{2009OtsukaPRB, 2010OtsukaPRB, 2012OtsukaPRB}.

Then we measured real-time tunneling into the probe SLQD.
Figure~\ref{FastMeas}(c) shows $V_{\rm rf}$ as a function of time $t$ with changing $V_{\rm PP}$ around point A in Fig.~\ref{Device}(b).
This real time measurement of charge transition is often used as a benchmark of fast electronic measurements~\cite{2004VandersypenAPL, 2007VinkAPL}.
$t_{\rm meas}$ was fixed at 5$\mu $s.
The traces (offset by 50~mV) show transitions between $V_{\rm PP}=$-1553.3 to -1551.5~mV.
As we increase $V_{\rm PP}$, the charge state changes from $n$ to $n+1$ and jumps of $V_{\rm rf}$ are observed around $V_{\rm PP}=$-1552.4~mV.
We could resolve tunneling events as fast as several tens of $\mu $s and this result is consistent with the result in Fig.~\ref{FastMeas}(b).
With less negative values of $V_{\rm PP}$, the state converges to the $n+1$ charge state.

\subsection{Measurement of local electronic states in a target quantum dot}
%%% Tunnel %%%

\begin{figure}
\begin{center}
  \includegraphics{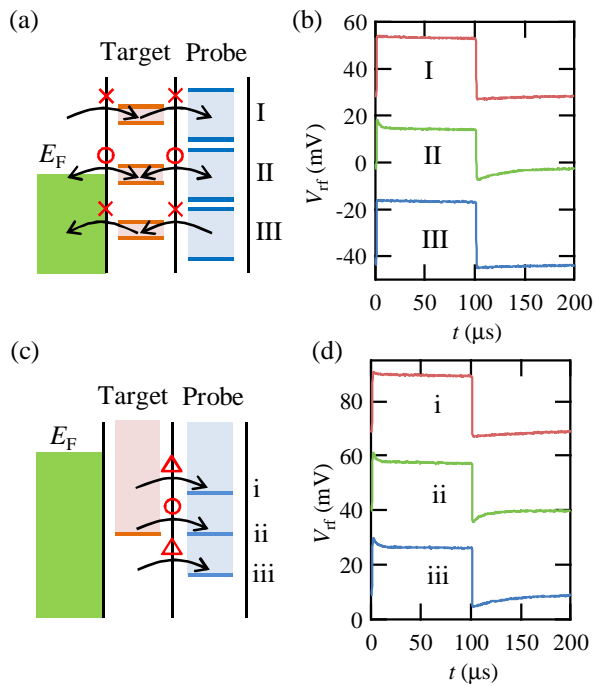}
  \caption{(color online) (a) Schematic of an energy diagram with applied square wave voltages on $V_{\rm PP}$.
  The levels in the probe SLQD and the target QD are shifted by the square wave (filled pairs of lines). 
  In case I and III, tunneling of electrons does not occur.
  Tunneling only occurs in case II.
  (b) $V_{\rm rf}$ as a function of $t$.
  The origin of the horizontal axis is set at falling edges of the square wave.
  The plotted value of $V_{\rm rf}$ is a result of averaging 4096 measurements.
  The traces (offset by -20~mV) show results with $V_{\rm PP}$=-1070, -1050, -1030~mV corresponding to case I, II and III, respectively.
  (c) Close up of the energy diagram in case II and in a phase of electron tunneling from the target QD to the probe SLQD.
  Tunneling is enhanced in case ii.
  (b) $V_{\rm rf}$ as a function of $t$.
  The traces (offset by -20~mV) show results with $V_{\rm PP}$=-1053, -1052, -1051~mV corresponding to case i, ii and iii, respectively.
  }
  \label{Tunnel}
\end{center}
\end{figure}

Next, we formed the target QD by applying negative voltages on the gates at the lower side of the device in Fig.~\ref{Device}(a).
We probed inner electronic states of the target QD by using the SLQD probe.
To detect the states of the target QD, we use the tunneling of electrons between the target QD and the probe SLQD.
We now apply a continuous square wave voltage on $V_{\rm PP}$ in order to periodically induce electron tunneling from the target QD to the probe SLQD.
The corresponding energy diagram is shown in Fig.~\ref{Tunnel}(a).
The levels in the probe SLQD and the target QD are shifted by the applied square waves and are shown as pairs of lines.
The filled ranges between the lines indicate windows in which the levels move.
When the Fermi level of the reservoir $E_{\rm F}$ is in the window of the target QD, and the window of the target QD is in the window of the probe SLQD as in case II, tunneling of electrons is synchronized with the applied square waves.
In the other cases, I and III, the tunneling does not occur because the levels are kept empty or filled in all phases of the square waves.

Figure~\ref{Tunnel}(b) shows measurement of electron tunneling into the probe SLQD in the time domain.
$V_{\rm rf}$ is plotted as a function of time $t$.
We applied a square wave with amplitude 16~mV and frequency 5~kHz.
The origin of the horizontal axis is set at falling edges of the square wave.
The plotted values of $V_{\rm rf}$ are the result of averaging 4096 measurements similar to measurements shown in Fig.~\ref{FastMeas}(c).
The traces (offset by -20~mV) show results with $V_{\rm PP}$=-1070, -1050, -1030~mV and these correspond to case I, II and III, respectively.
In cases I and III, there is no electron tunneling and $V_{\rm rf}$ exhibits a square wave shape resulting from the direct electrostatic coupling between the applied square wave voltage and the charge sensor.
On the other hand, in case II, electron tunneling events are synchronized with the square wave.
This tunneling process is stochastic and produces exponential-decay changes of $V_{\rm rf}$ on the square wave background by averaging a large number of time traces.

Figure~\ref{Tunnel}(c) is a close-up of the energy diagram for case II and in a case of electron tunneling from the target QD to the probe SLQD.
If the level of the target QD is not aligned to the level of the probe SLQD like in cases i and iii, tunneling is suppressed as it is an inelastic process.
On the other hand, when the levels are aligned as in case ii, tunneling is an elastic process and enhanced~\cite{1995vanderVaartPRL, 1998FujisawaScience}.
Therefore, we can detect the target level as an enhancement of the tunneling into the probe SLQD.
(The same mechanism also works in a case of tunneling from the probe SLQD to the target QD.)

The corresponding data in such a scheme is shown in Fig.~\ref{Tunnel}(d).
Traces (offset by -20~mV) show the results with $V_{\rm PP}$=-1053, -1052, -1051~mV and these correspond to case i, ii and iii, respectively.
The average electron tunneling time from the target QD to the probe SLQD in case ii (12~$\mu $s) is shorter than the values in cases i (105~$\mu $s) and iii (39~$\mu $s) as expected.
This result shows that operation of the detection scheme utilizing tunneling of electrons is confirmed by the measurement of electron tunneling with the wide-band probe.

%%% Measurement of QD state %%%

\begin{figure}
\begin{center}
  \includegraphics{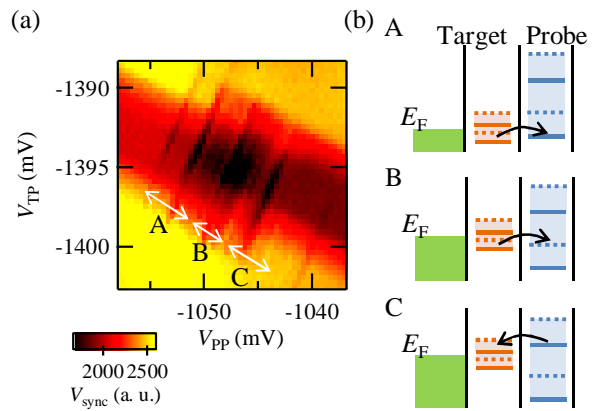}
  \caption{(color online) (a) $V_{\rm sync}$ as a function of $V_{\rm PP}$ and $V_{\rm TP}$.
  Black lines show conditions at which the levels in the target QD and the probe SLQD align and electron tunneling is enhanced.
  (b) Energy diagrams in regions A, B and C.
  The solid and dotted lines show the ground and the excited states, respectively.
  In regions A, B and C, tunneling occurs from the states in the target QD to the ground state in the SLQD, from the states in the target QD to the excited state in the SLQD, and from the ground state of the SLQD to the states in the target QD.
  }
  \label{QDState}
\end{center}
\end{figure}

Next, we checked the modulation of the tunneling with changing $V_{\rm PP}$ and a plunger gate voltage of the target QD $V_{\rm TP}$.
Figure~\ref{QDState}(a) shows a synchronized component of $V_{\rm rf}$ with the applied square wave $V_{\rm sync}$ measured by using the lock-in technique as a function of $V_{\rm PP}$ and $V_{\rm TP}$.
$V_{\rm sync}$ will be decreased when electron tunneling happens as shown in Fig.~\ref{Tunnel}(d)ii.
We can see several black lines from the lower left to the upper right and these correspond to conditions at which the levels in the target QD and the probe SLQD align and electron tunneling occurs.
By considering the energies of excited states in both QDs~\cite{comment3}, lines in region A correspond to electron tunneling from states in the target QD to the ground state in the SLQD, region B corresponds to tunneling from states in the target QD to the excited state in the SLQD, and region C corresponds to tunneling from the ground state of the probe SLQD to states in the target QD as shown in Fig.~\ref{QDState}(b).
The solid and dotted lines in the figure show the ground and the excited states, respectively.

The thick band structure from the upper left to the lower right corresponds to the electron tunneling between the target QD and the reservoirs because the charge sensor also has finite sensitivity to the charges in the target QD.
The relative signal intensity by the target QD is about half of that by the probe QD reflecting the smaller capacitive coupling between the target QD and the sensor.
The lower left edges of the black lines are points at which the target QD level is aligned to the Fermi level of the reservoirs in the injection phase.
Moving to the upper right along the black lines, the target level goes below the Fermi level.
The upper right edges of the black lines correspond to the points at which the target level is 450~$\mu $eV below the Fermi level.
Even at this condition, we could detect the target levels with the SLQD probe.
We observe electronic states of the target QD with line width as small as 30~$\mu $eV by considering that the coupling to the probe SLQD (several tens of $\mu $s) does not broaden the target levels.
These results show that the SLQD probe can access levels, not just those around the Fermi level, with good energy resolution.

\subsection{Robustness of the measurement by a single-lead quantum dot probe}
%%% Robustness of the measurement %%%

\begin{figure}
\begin{center}
  \includegraphics{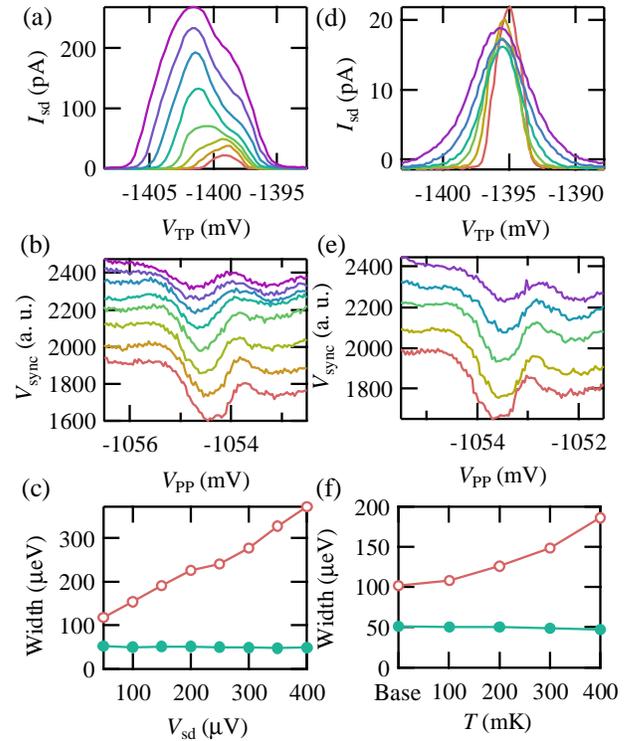}
  \caption{(color online) (a), (b) Bias voltage dependence of Coulomb peaks through the target QD (a) and tunneling signal measured by the SLQD probe (b).
  $V_{\rm sd}$ is changed from 50 to 400~$\mu $V with 50~$\mu $V step, from the bottom to the top.
  Offsets are set as $\Delta V_{\rm sync}$=50 in Fig.~(b).
  (c) Evaluated signal width as a function of $V_{\rm sd}$.
  The open (filled) circles show results in Coulomb peaks (tunneling signal by the SLQD probe).
  (d), (e) Temperature dependence of Coulomb peaks through the target QD (d) and tunneling signal measured by the SLQD probe (e).
  $T$ is changed from the base temperature to 400~mK with 100~mK step, from the bottom to the top.
  Offsets are set as $\Delta V_{\rm sync}$=50 in Fig.~(e).
  (c) Evaluated signal width as a function of $T$.
  The open (filled) circles show results in Coulomb peaks (tunneling signal by the SLQD probe).
  }
  \label{Robustness}
\end{center}
\end{figure}

Next, we checked the robustness of our measurement scheme against the electronic distribution in the reservoirs of the device.
Figures~\ref{Robustness}(a) and (b) show bias voltage dependence of Coulomb peaks through the target QD (a) and the corresponding tunneling signal measured by the SLQD probe (b).
The traces (offset by 50 in Fig.~\ref{Robustness}(b)) show results when we change the source drain bias voltage of the target QD $V_{\rm sd}$ from 50 to 400~$\mu $V with 50~$\mu $V step.
The width of the Coulomb peaks are shown as open circles in Fig.~\ref{Robustness}(c).
The width is taken at 1/3 of the peak height in order to prevent the effect of excited states, which changes the shape of the peaks~\cite{1992JohnsonPRL}.
We observe an increase of the width with the increase of $V_{\rm sd}$.
Filled circles in Fig.~\ref{Robustness}(c) are the width of the SLQD probe signal evaluated by FWHM.
Strikingly, the width of the peaks is unaffected by the change of $V_{\rm sd}$. 
This proves that the measurement by the SLQD probe is robust against $V_{\rm sd}$.

Then, we studied the temperature dependence of the Coulomb peaks and the SLQD probe signal, as shown in Fig.~\ref{Robustness}(d) and (e).
The traces (offset by 50 in Fig.~\ref{Robustness}(e)) show results with changing temperature of the fridge $T$ from base temperature (13~mK) to 400~mK with 100~mK steps.
The FWHM of the Coulomb peaks as a function of $T$ is shown as open circles in Fig.~\ref{Robustness}(f).
The width increases with the increase of $T$ reflecting the broadening of the Fermi distribution in the reservoirs. 
The width of the SLQD probe signal, shown as filled circles in fig 5(f), shows no dependence on $T$.
These results show that the measurement by the SLQD probe can access the local electronic states precisely, liberated from the electronic distribution of the reservoirs.

\subsection{Real-time measurement of the target quantum dot states}
%%% Real-time measurement of the target QD %%%

\begin{figure}
\begin{center}
  \includegraphics{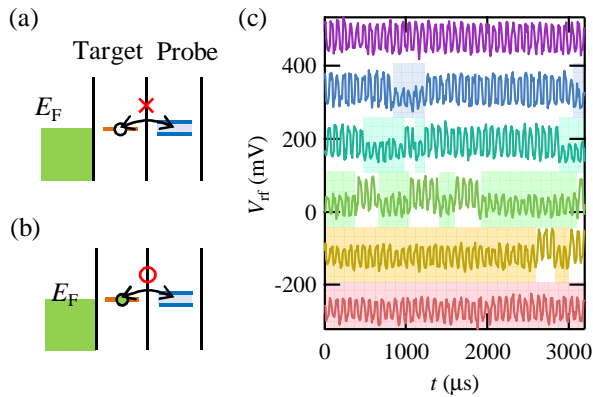}
  \caption{(color online) (a), (b) Energy diagram when small square wave voltages are applied on $V_{\rm PP}$.
  If the target QD is empty (filled), inter-dot tunneling does not occur (a) (does occur (b)).
  (c) Real-time detection of charges in the target QD. 
  $V_{\rm rf}$ is plotted as a function of $t$.
  The traces show results when QD levels are changed against the Fermi energy.
  In the shaded regions, the amplitude becomes small because the target QD is filled and inter-dot tunneling occurs.
  The tunneling electrons screen the square wave and the amplitude of $V_{\rm rf}$ decreases.
  }
  \label{Realtime}
\end{center}
\end{figure}

Finally, we monitored the charge state of the target QD by using the fast SLQD probe.
To detect the charge states of the target QD in real time with the SLQD probe, we again use the inter-dot tunneling when we apply small a square wave voltage on $V_{\rm PP}$.
Measurements and refreshments of the probe are repeated by applying the square wave.
We lowered the response time of the SLQD probe by making the inter-dot tunneling faster than 5~$\mu $s and also by making the charge transition time of the target QD slower in the accessible range of the probe.
If the target QD level is empty, inter-dot tunneling does not occur (Fig.~\ref{Realtime}(a)).
On the other hand, if the level is filled, inter-dot tunneling synchronized with the square wave occurs (Fig.~\ref{Realtime}(b)).
This inter-dot tunneling can be detected by the SLQD probe.

Figure~\ref{Realtime}(c) shows the observed $V_{\rm rf}$ as a function of time.
We applied a square wave (frequency 12.5~kHz, amplitude 2~mV) and the resulting square wave shaped $V_{\rm rf}$, which comes from direct electrostatic coupling between $V_{\rm PP}$ and the charge sensor, is observed.
The traces show the results when we changed the target QD levels from beneath the Fermi level to above.
We can observe two kinds of oscillation amplitudes.
If an inter-dot electron tunneling occurs, the electron screens the applied square wave and the amplitude decreases.
Therefore, the region with the small amplitude shaded in Fig.~\ref{Realtime}(c) corresponds to the filled condition of the target QD.
We can clearly see the real-time jump of the target QD charge sate and the jump events depend on the energy of the target QD levels against the Fermi level.

\section{Discussion}

In conclusion, we have realized a fast SLQD probe, which can access local electronic states in nanostructures with wide bandwidth.
We evaluated the operation speed of the probe and applied this new probe to a measurement of the states in a target QD.
We confirmed the operation of the probe and demonstrated characteristic properties of the probe. 

The new SLQD probe shows three superior properties compared to previous probes; (1) the SLQD probe can access more states, not just those around the Fermi level, (2) measurement by the SLQD probe is robust against change of electronic distribution in the reservoirs, (3) fast real-time measurement is possible.
Properties (1) and (2) are the result of the single-lead quantum dot structure, in which the probe state is fully isolated and freed from electronic distribution of the electron bath.
These properties will be powerful for measuring fragile local electronic states and their dynamics in nanostructures.
For example, the probe will be useful to probe Kondo states under non-equilibrium conditions~\cite{1993MeirPRL, 2002DeFranceschiPRL, 2005LeturcqPRL} and to readout the electron spin in QDs~\cite{2004ElzermanNature, 2011NowackScience} with high fidelity even at relatively high electron temperatures.

\section{Methods}
\subsection{Device structure and measurement}
The device was fabricated from a GaAs/AlGaAs heterostructure wafer with sheet carrier density 2.0~$\times$~10$^{15}$~m$^{-2}$ and mobility 110~m$^2$/Vs at 4.2~K.
The two-dimensional electron gas is formed 90~nm beneath the surface.
We patterned a mesa structure by wet-etching and formed Ti/Au Schottky surface gates by metal deposition, which appear white in Fig.~\ref{Device}(a).
All of the measurements were conducted in a dilution fridge at a temperature of 13~mK.

\section{Acknowledgements}
We thank J. Beil, J. Medford, F. Kuemmeth, C. M. Marcus, D. J. Reilly, K. Ono, RIKEN CEMS Emergent Matter Science Research Support Team and Microwave Research Group in Caltech for fruitful discussions and technical supports.
Part of this work is supported by the Grant-in-Aid for Research Young Scientists B, Funding Program for World-Leading Innovative R\&D on Science and Technology (FIRST) from the Japan Society for the Promotion of Science, ImPACT Program of Council for Science, Technology and
Innovation, Toyota Physical \& Chemical Research Institute Scholars, RIKEN Incentive Research Project, Yazaki Memorial Foundation for Science and Technology Research Grant, Japan Prize Foundation Research Grant, Advanced Technology Institute Research Grant, IARPA project ``Multi-Qubit Coherent Operations'' through Copenhagen University, Mercur  Pr-2013-0001, DFG-TRR160,  BMBF - Q.com-H  16KIS0109, and the DFH/UFA  CDFA-05-06.

\section{Author contributions}
T. O. and S. T. planned the project; T. O., S. A., A. L. and A. W. performed device fabrication; T. O., S. A. T. N., M. D., J. Y., K. T., R. S., G. A. and S. T. conducted experiments and data analysis; all authors discussed the results; T. O., S. A., T. N., M. D., G. A. and S. T. wrote the manuscript.


\begin{references}
\bibitem{comment1}
In experiments in the range of this paper, $n=1$ is not necessary.
\bibitem{comment3}
Excited orbital energies of the both QDs are evaluated in another experiment.
The values are 200 and 480~$\mu $eV for the target QD and the probe SLQD, respectively.
\end{references}
\end{document}